\documentclass[12pt,a4paper]{article}
\usepackage{amsmath,amsfonts,amssymb,amsthm}
\usepackage{cite}
\usepackage{times}
\usepackage{pdfsync}
\usepackage{hyperref}

\newcommand{\be}{\begin{equation}}
\newcommand{\ee}{\end{equation}}
\newcommand{\bea}{\begin{eqnarray}}
\newcommand{\eea}{\end{eqnarray}}

\def\d_Vphi{\text{d}_V\hspace{-0.06em}\phi}
\def\d_Vphibar{\text{d}_V\hspace{-0.06em}\bar\phi}
\def\d_Vxi{\text{d}_V\hspace{-0.06em}\xi}

\def\be{\begin{eqnarray}}
\def\ee{\end{eqnarray}}
\def\beann{\begin{eqnarray*}}
\def\eeann{\end{eqnarray*}}
\def\beq{\begin{equation}}
\def\eeq{\end{equation}}
\def\ba{\begin{array}}
\def\ea{\end{array}}
\def\ben{\begin{enumerate}}
\def\een{\end{enumerate}}
\def\bea{\begin{eqnarray}}
\def\eea{\end{eqnarray}}

\def\5{\bar }
\def\6{\partial }
\def\7{\hat }
\def\4{\tilde }
\advance\textheight\topskip
\renewcommand{\tilde}{\widetilde}
\renewcommand{\hat}{\widehat}

\renewcommand{\d}{\partial}

\newcommand{\binner}[2]{%
  {\langle}\kern-4.15pt{\langle}#1{,}\,#2{\rangle}\kern-4.15pt{\rangle}}

\newcommand{\half}{\frac{1}{2}}

\newcommand{\ffrac}[2]{\raisebox{.5pt}%
  {\footnotesize$\displaystyle\frac{#1}{#2}$}\kern1pt}

\def\cH{\mathcal{H}}

\def\cL{\mathcal{L}}

\def\cP{\mathcal{P}}

\numberwithin{equation}{section} \makeatletter
\@addtoreset{equation}{section}

\DeclareFontFamily{OT1}{rsfs}{} \DeclareFontShape{OT1}{rsfs}{m}{n}{
<-7> rsfs5 <7-10> rsfs7 <10-> rsfs10}{}
\DeclareMathAlphabet{\mycal}{OT1}{rsfs}{m}{n}

\begin{document}

\title{Black hole entropy from non-proper gauge degrees of freedom:
  II. The charged vacuum capacitor}

\author{Glenn Barnich}

\date{}

\def\mytitle{Black hole entropy from non-proper gauge degrees of
  freedom: II. The charged vacuum capacitor}

\pagestyle{myheadings} \markboth{\textsc{\small G.~Barnich}}{%
  \textsc{\small Quantum mechanics of non-proper gauge degrees of
    freedom}}

\addtolength{\headsep}{4pt}


\begin{centering}

  \vspace{1cm}

  \textbf{\Large{\mytitle}}


  \vspace{1.5cm}

  {\large Glenn Barnich}

\vspace{.5cm}

\begin{minipage}{.9\textwidth}\small \it  \begin{center}
   Physique Th\'eorique et Math\'ematique \\ Universit\'e libre de
   Bruxelles and International Solvay Institutes \\ Campus
   Plaine C.P. 231, B-1050 Bruxelles, Belgium
 \end{center}
\end{minipage}

\end{centering}

\vspace{1cm}

\begin{center}
  \begin{minipage}{.9\textwidth}
    \textsc{Abstract}. The question which degrees of freedom are
    responsible for the semi-classical contribution to the partition
    function, obtained by evaluating the Euclidean action improved
    through suitable boundary terms, is addressed. A physical toy
    model for the gravitational problem is a charged vacuum
    capacitor. In Maxwell's theory, the gauge sector including ghosts
    is a topological field theory. When computing the grand canonical
    partition function with a chemical potential for electric charge
    in the indefinite metric Hilbert space of the BRST quantized
    theory, the classical contribution to the partition function
    originates from the part of the gauge sector that is no longer
    trivial due to the boundary conditions required by the physical
    set-up. More concretely, for a planar charged vacuum capacitor
    with perfectly conducting plates, we identify the degrees of
    freedom that, in the quantum theory, give rise to additional
    contributions to the standard black body result proportional to
    the area of the plates, and that allow for a microscopic
    derivation of the thermodynamics of the charged capacitor.
  \end{minipage}
\end{center}

\vfill

\thispagestyle{empty}
\newpage

\begin{small}
{\addtolength{\parskip}{-2pt}
 \tableofcontents}
\end{small}
\thispagestyle{empty}
\newpage

\section{Introduction}
\label{sec:introduction}

The question which degrees of freedom are responsible for the
Bekenstein-Hawking entropy of black holes naturally leads one to study
non-proper gauge degrees of freedom, i.e., gauge degrees of freedom
that are no longer pure gauge because of non trivial boundary
conditions. (i) The most direct line of reasoning is probably to
consider the Hamiltonian formulation of linearized Einstein
gravity. The linearized Schwarzschild solution does not involve
physical degrees of freedom since the transverse-traceless parts of
the spatial metric and its momenta vanish for that solution. (ii)
Another argument, which holds on the non-linear level, concerns the
Bekenstein-Hawking entropy of the black hole in three dimensional
anti-de Sitter spacetime where there are no physical bulk gravitons to
begin with. (iii) Yet another approach has to do with the type of
observables that are involved: in general relativity, the ADM mass is
a codimension 2 surface integral, with similar properties to electric
charge in Maxwell's theory. In particular, it does not involve
transverse-traceless variables. Furthermore, the classification of
such observables is directly related to non-proper diffeomorphisms or
large gauge transformations.

One possibility is to introduce the non-trivial boundary
conditions as dynamical canonical variables in the theory, with
suitable additional constraints. This idea goes back to Dirac
\cite{dirac:1964xx} and has been used in an investigation of the
definition of energy, and more generally of the Poincar\'e generators,
in the Hamiltonian formulation of asymptotically flat general
relativity \cite{Regge:1974zd}. In the context of Yang-Mills theory,
it has been implemented for various related questions
\cite{Gervais:1976ec,Benguria:1976in,Wadia:1976fa,Wadia1977,%
  Gervais:1978kn,Wadia1979}, including the infrared problem
\cite{Gervais:1980bz}. 

These arguments suggest to study the analogue problem in the
context of the quantized electromagnetic field, where the role of the
black hole is played by the Coulomb solution, the electromagnetic
field created by a static point particle source with macroscopic
charge $Q$. Besides being a physical problem in its own right where
all conceptual issues are present, the linearity of the problem and
the wealth of results readily available in the literature make it
directly tractable.

In the first paper of this series \cite{Barnich:2010bu}, a quantum
mechanical understanding has been achieved when all polarizations of
the photon are quantized in an indefinite metric Hilbert space: the
quantum state $|0\rangle^Q$ corresponding to the classical Coulomb
solution is a coherent state of null oscillators, made up of a
linear combination of longitudinal and temporal photons. In this
computation, infrared divergences occur when showing that the
expectation value ${}^Q\langle 0|\hat \pi^i(x)|0\rangle^Q$ of the
electric field operator is indeed the classical field produced by a
point-like source: one uses that the Fourier transform of $k^{-2}$ is
proportional to $(4\pi r)^{-1}$ which really requires an infrared
regularisation, $(k^2+m^2)^{-1}$ giving the Yukawa potential
$(4\pi r)^{-1}e^{-m r}$, with $m\to 0^+$.

Unlike ordinary coherent states, null coherent states have the same
norm than the standard vacuum, $^Q\langle 0|0\rangle
^Q=1$. Furthermore, the expectation value of the energy of physical
photons vanishes. It is in this sense that these states behave like
different vacua of the theory.

That longitudinal and temporal photons have an important role to play
in topologically non-trivial situations is in agreement with the
standard interpretation of the Aharanov-Bohm effect
\cite{AharonovBohm1959} when extrapolated to the quantized
electromagnetic field.

Rather than quantizing the theory for a fixed charge, what we would
like to address here is the computation of the grand canonical
partition function,
\begin{eqnarray}
  Z(\beta,\mu)={\rm Tr}\, e^{-\beta(\hat H
    -\mu \hat Q)}, \label{eq:1}
\end{eqnarray}
with a precise understanding of the underlying Hilbert space and thus
of the trace that is involved. Again, when trying to deal directly
with the electric charge operator, 
\begin{equation}
\hat Q=-\int_{\partial {\cal V}}
d\sigma_i\, \hat\pi^i=-\int_{\cal V} d^3x\, \partial_i \hat
\pi^i,\label{eq:10} 
\end{equation}
in a large volume ${\cal V}$, one has to face infrared questions since
$-\hat Q$ is the zero mode of the longitudinal part of the electric
field. 

On the classical level, the role of the chemical potential is played
by the constant value of $A_0=-\mu$ at the surface of the body, while
a non-vanishing electric charge requires $\pi^r=O(r^{-2})$. In order
to take electric charge into account, non trivial fall-off or boundary
conditions are thus required.

The approach we will follow here is not to introduce additional
degrees of freedom besides those already contained in
$(A_\mu,\pi^\mu)$. For trivial boundary conditions, standard results
equivalent to those derived in the framework of reduced phase space
quantization are then recovered in the indefinite metric BRST Fock
space through the quartet mechanism \cite{Kugo:1979gm} in the bulk. We
will analyze in detail how these results are modified when imposing
the boundary conditions that are used in the context of the Casimir
effect \cite{casimir1948attraction}. For technical reasons, it is then
also easier for us here to start with a vacuum capacitor consisting of
2 large parallel plates instead of a spherical vacuum capacitor, so
that one may use Fourier series instead of Bessel functions
\cite{Boyer:1968uf}.

Recent work on infrared physics has been driven by new connections in
the field summarized in \cite{Strominger:2017zoo}. There is a
considerable overlap of ideas and results underlying this computation
here and those developed in terms of edge modes in
\cite{Balachandran:1994vi,Kabat:1994vj,Kabat:2010nm,Donnelly:2014fua,%
  Donnelly:2015hxa,Donnelly:2016auv,Seraj:2016jxi,Geiller:2017xad,%
  Geiller:2017whh,Seraj:2017rzw,Blommaert:2018rsf,Blommaert:2018oue}. A
detailed comparision, also with the considerations in
\cite{Henneaux:2018gfi}, deserves further investigation.

The paper is organized as follows. In the next section, we start by
discussing the thermodynamics of a charged vacuum capacitor following
the method developed by Gibbons and Hawking \cite{Gibbons:1976ue}:
from the Euclidean path integral, it follows that the semi-classical
approximation to $\ln Z(\beta,\mu)$ is given by minus the Euclidean
action evaluated at the classical solution. The appropriate boundary
terms needed for the charged capacitor have already been introduced in
the context of charged black holes for instance in
\cite{Hawking:1995ap}. As compared to the one-loop result for the
standard black body, there is now a contribution proportional to the
area coming
from the classical saddle point, together with additional
contributions at one-loop. The
purpose of this paper is to provide a microscopic derivation of the
saddle point and the additional contributions to the partition
function. 

In section \ref{sec:topologicalFT}, we point out in what sense the
gauge sector of Maxwell's theory can be understood as a topological
field theory. It is not really needed for the rest of the paper, but
is included in order to better understand the relation with three
dimensional gravity for instance.

In section \ref{sec:plan-vacu-capac}, boundary conditions adapted to
perfectly conducting parallel plates, taken at constant $z$, are
imposed. Through a detailed Hamiltonian analysis, we show that the
modes with vanishing momenta in the $z$ direction of $(A_z,\pi^z)$,
even though formally longitudinal, are to be considered as physical in
the problem at hand. In that sense, we refer to them as non-proper
gauge degrees of freedom.

In the quantum theory, we compute in section
\ref{sec:partition-function} the contribution of the non-zero modes of
the non-proper gauge degrees of freedom to the standard black body
result. It is proportional to the area of the plates. After turning on
the chemical potential for electric charge, a quantum mechanical
understanding of the classical thermodynamics of the vacuum capacitor
follows from the contribution of the zero mode of the non-proper gauge
degrees of freedom.

Additional remarks are relegated to section
\ref{sec:disc-persp}. Conventions for mode expansions adapted to the
various boundary conditions are given in appendix \ref{sec:modes}. In
order to be self-contained, a summary of standard material on BRST
quantization as applied to Maxwell's theory is provided in appendix
\ref{sec:trivial} and appendix \ref{sec:coher-stat-quart}.

\section{Thermodynamics of a charged vacuum capacitor} 
\label{sec:classical-thermo}

When making the Legendre transformation of the standard Lagrangian action
$S[A_\mu]=-\frac{1}{4}\int d^4x\, F_{\mu\nu} F^{\mu\nu}$ for $\dot
A_i$, and after adding  the boundary term, 
$-\oint_{\partial B} d\sigma_i [\pi^i A_0]$, the first order action
is  
\begin{equation}
  \label{eq:9}
  I=\int d^4x [\dot A_i \pi^i - \cH_0 +A_0\d_i\pi^i+j^\mu A_\mu],
  \quad \cH_0=\half
  (\pi^i\pi_i + B^iB_i), 
\end{equation}
where $B^i=\epsilon^{ijk}\d_jA_k$, $E^i=-\pi^i$. Alternatively, this
action may be obtained from the extended first order action after
eliminating the Lagrange multiplier for the primary constraint and the
momentum $\pi^0$.

From the viewpoint of constrained Hamiltonian systems, there are two
gauge invariant observables in the problem, the reduced phase space
energy
\begin{equation}
  \label{eq:23}
  H^{\rm ph}=\int d^3x\, \cH^{\rm ph},\quad \cH^{\rm ph}=
  \half (\pi^i_T \pi_i^T-A_i^T\Delta A^i_T), 
\end{equation} 
and also the electric charge
\begin{equation}
  \label{eq:27}
  Q=-\int_S d\sigma_i\ \pi^i_L, 
\end{equation}
where $S$ is a closed 2-surface.

Consider a spherical vacuum capacitor consisting of two conducting
spheres $S_1,S_2$ centered at the origin with radii $R_1 < R_2$ and
charges $q$, $-q$. Let us first focus on time-independent fields and
assume that there are no sources inside the body. We will assume here
that $A_i=0$, even though the field equations only require
$\d_j F^{ji}=0$. In this context, there are then no transverse degrees
of freedom and
\begin{equation}
  A_0=-\phi=-\frac{q}{4\pi r},\quad \pi^i=-\frac{q x^i}{4\pi
    r^3} \label{eq:28} 
\end{equation}
for $R_1 < r < R_2$ and zero otherwise.

The thermodynamics can then be obtained from the Euclidean action
evaluated on-shell. Since the problem is at fixed electric charge, no
improvement boundary terms are needed \cite{Deser:1997xu}, and
\begin{equation}
  I_{E}=\frac{\beta}{2}\int d^3x\
  \pi^i_L\pi^L_i=\half c_S \beta q^2,\quad
  c_S=\frac{R_2-R_1}{4\pi R_1R_2}. \label{eq:29} 
\end{equation} 
 Using
$\pi^i_L=\d^i\phi$ and $\Delta \phi=0$ on-shell for $R_1<r<R_2$,
$\int d^3x\ \pi^i_L\pi_i^L= \int d^3x\,\d_i (\phi \d^i\phi)$,
$I_E$ can also be written in terms of boundary
terms as
\begin{equation}
  \label{eq:30}
  I_E=-\frac{\beta}{2}(\phi|_{S_2}-\phi|_{S_1})Q,  
\end{equation}
where $\phi_S=\frac{q}{4\pi r}$ and $Q=q$ for the problem at
hand. This then gives rise to the semi-classical contribution to the
partition function,
\begin{equation}
  \label{eq:31}
  \ln Z(\beta,Q)=-I_E(\beta,Q)+f(\beta), 
\end{equation}
where one would expect $f(\beta)$ to be given by the standard one-loop
contribution of physical photons,
\begin{equation}
f_V(\beta)=\frac{1}{3} b_V\beta^{-3},\quad b_V=\frac{\pi^2V}{15}. \label{eq:33}
\end{equation}
The analysis below shows however that there are additional contributions
\begin{equation}
  \label{eq:100}
  f(\beta)=f_V(\beta)+f_A(\beta)-\half \ln (2\pi \beta),\quad f_A(\beta)=\frac{1}{2}b_A\beta^{-2},
\end{equation}
with $b_A$ proportional to the area, 
\begin{equation}
b_A=\frac{\zeta(3)}{\pi}A,\label{eq:101}
\end{equation}
in the case of the planar capacitor\footnote{The conditions under
  which some of these terms can be neglected will be discussed
  elsewhere.}.  This implies that
\begin{equation}
  \label{eq:97}
  U=-\frac{\d\ln Z(\beta,
    Q)}{\d\beta}=-f'(\beta)+\frac{1}{2}c_S Q^2. 
\end{equation}
In case this can be inverted to yield $\beta=\beta(U')$, with
$U'=U-\half c_S Q^2$, 
the entropy is 
\begin{equation}
  \label{eq:98}
  S(U,Q)=(1-\beta\frac{\d}{\d\beta}) f(\beta)|_{\beta=\beta(U')}. 
\end{equation}

Alternatively, in order to deal directly with 
\begin{equation}
Z(\beta,\mu)={\rm Tr}\ e^{-\beta(\hat H-\mu \hat Q)},\label{eq:34}
\end{equation}
one supposes instead that the electric potentials at the boundary are
fixed and constant, $\phi|_{S_1}=\phi_1$, $\phi|_{S_2}=\phi_2$ with
$\mu=\phi_1-\phi_2$. Under the additional assumptions that there are
no sources inside the body, $\d^i A_i=0$ and $A^T_i=0=\pi^i_T$, the
classical solution is
\begin{equation}
  \phi=\frac{1}{R_2-R_1}(R_2\phi_2-R_1\phi_1+
  \frac{\mu R_1R_2}{r}),\quad \label{eq:35}
E^i=\frac{\mu R_1R_2\,x^i}{(R_2-R_1)r^3}.  
\end{equation}
In this situation, following \cite{Regge:1974zd} (see also
\cite{Hawking:1995ap}), the action needs to be improved by boundary
terms so that this solution is a true extremum of the variational
principle,
\begin{equation}
  \label{eq:36}
  I'=I+\int dt \phi_2 Q-\int dt \phi_1 Q
\end{equation}
On-shell, the Euclidean action is now
\begin{equation}
  \label{eq:37}
  I'_E=\frac{\beta}{2}(\phi_2-\phi_1)Q, \quad Q=c_S^{-1}\mu.  
\end{equation}
This leads to
\begin{equation}
  \label{eq:38}
  I'_E=-\frac{1}{2} c_S^{-1}\beta\mu^2. 
\end{equation}
Together with the one-loop results, one thus finds 
\begin{equation}
  \label{eq:39}
  \ln Z(\beta,\mu)=-I'_E+f(\beta).
\end{equation}
The electric charge is then 
\begin{equation}
Q=\beta^{-1}\frac{\partial\ln Z(\beta,\mu)}{\partial\mu}=c_S^{-1}\mu.\label{eq:99}
\end{equation}
At fixed $\beta$, the Legendre transform of $\ln Z(\beta,\mu)$ with
respect to $\mu$, 
\begin{equation}
\ln Z(\beta,Q)=(1-\mu\frac{\partial}{\partial\mu})\ln
Z(\beta,\mu)|_{\mu=\mu(Q)}, \label{eq:102}
\end{equation}
then leads back to \eqref{eq:31}.

For the case of the so-called exterior problem, the thermodynamics of a charged
spherical shell of radius $R_1$ can be obtained from the above by
letting $R_2\to\infty$ and taking $\phi_2=0$.

For two parallel plates $P_1,P_2$ at $z=0$ and at $z=L_3$, with charge
densities $\frac{q}{A}$ and -$\frac{q}{A}$, one finds under the same
assumptions and in the same manner that
$\pi^i=-\delta^i_3\frac{q}{A}$, (when $x^i=(x,y,z)$),
$\phi=-\frac{q}{A}z$, with $\mu=\frac{L_3q}{A}$. The only change in
the classical part of the above discussion is then the replacement of
the geometric factor $c_S$ by
\begin{equation}
  \label{eq:32}
  c_P=\frac{L_3}{A}. 
\end{equation}

What we will study below is the quantum mechanical origin of
the semi-classical contribution to the partition function, together
with the additional one-loop contributions.

\section{Gauge sector of electromagnetism as topological
  field theory}
\label{sec:topologicalFT}

The gauge sector including ghosts of Maxwell's theory is treated in
the context of the Batalin-Fradkin-Vilkovisky Hamiltonian formalism
\cite{Fradkin:1975cq,Batalin:1977pb,Fradkin:1978xi}. It contains the
information on the electric charge in regions where there are no
sources. Not surprisingly, this sector can be identified with a
Witten-type supersymmetric quantum mechanical model \cite{Witten1982}
when treating the spatial dimensions in a formal way. We follow the
reviews \cite{Henneaux:1992ig}, chapter 19, and
\cite{Birmingham:1991yq} chapter 3, for the BFV treatment of
electromagnetism and for supersymmetric quantum mechanics,
respectively.

In the non-minimal BFV-BRST approach in which $(A_0,\pi^0)$ are among
the canonical variables, the action to be used in the Hamiltonian path integral for
electromagnetism is
\begin{equation}
 S=\int dt\int d^3x \Big[\dot A_\mu \pi^\mu+\dot \eta {\cal
   P}+\dot{\bar C}\rho-\cH_0-\{\Omega,K_\xi\}\Big], 
\end{equation}
where the BRST invariant Hamiltonian is $H_0=\int d^3x\, \cH_0$,
$\cH_0$ is given in \eqref{eq:9}, and the graded Poisson brackets are
determined by
\begin{equation}
\{A_\mu(\vec x),\pi^\nu(\vec y)\}=\delta^\nu_\mu\delta^{(3)}(\vec
x,\vec y),\quad
\{\eta(\vec x),\cP(\vec y)\}=-\delta^{(3)}(\vec x,\vec y)=\{\bar
C(\vec x),\rho(\vec y)\}.\label{eq:53}
\end{equation}
The BRST charge is
\begin{equation}
\Omega=-\int d^3x (i\rho \pi^0 +\eta \partial_i \pi^i)
\label{eq:2}, 
\end{equation}
and the gauge fixing fermion is chosen as 
\begin{equation}
  K_\xi=-\int d^3x (i\bar C \partial_k A^k +{\cal P} A_0-\xi
\frac{i}{2}\bar C\pi^0),
\end{equation}
so that 
\begin{equation}
\{\Omega,K_\xi\}=\int d^3x (\partial_k A^k \pi^0-\partial_i \pi^i A_0
+i {\cal P}\rho +i \partial^i \bar C\partial_i
\eta-\frac{\xi}{2}\pi^0\pi^0).\label{eq:3}
\end{equation}
Eliminating the auxiliary fields $\pi^i\approx F^{i0}$,
$\pi^0\approx \frac{1}{\xi}(\d_\mu A^\mu)$, $\rho\approx i\dot\eta$,
$\cP\approx -i\dot{\bar C}$, gives the covariant gauge fixed Faddeev-Popov
action,
\begin{equation}
  \label{eq:21}
  S=\int d^4x\,[-\frac{1}{4}F^{\mu\nu}F_{\mu\nu}-\frac{1}{2\xi}(\d^\mu
  A_\mu)(\d^\nu A_\nu)-i\d^\mu \bar C\d_\mu \eta],
\end{equation}
but we will not do so here in order to keep better track of the
various degrees of freedom. 

Decomposing into transverse and longitudinal fields, $A_i=A_i^T+\d_i
A$, with $A=\frac{\d^i A_i}{\Delta}$,
$\pi^i=\pi ^i_T+\frac{1}{\Delta}\d^i\pi$ with $\pi=\partial_i \pi^i$,
the first order action decomposes into a transverse piece,
\begin{equation}
  \label{eq:14}
  S^{\rm ph}=\int dt\int d^3x (\dot A^T_i
  \pi^i_T-\cH^{\rm ph}),
\end{equation}
with $\cH^{\rm ph}$ given in \eqref{eq:23}, and a piece from the gauge
sector (including ghosts),
\begin{equation}
  \label{eq:17}
  S^{\rm gs}=\int dt\int d^3x \Big(\dot A_0 \pi^0-\dot A\pi+\dot \eta {\cal
    P}+\dot{\bar C}\rho - \cH^{\rm gs}),
\end{equation}
where 
\begin{equation}
  H^{\rm gs}=\int d^3x\, \cH^{\rm gs}=-\half i\{\Omega,\bar\Omega\},\quad
  \bar\Omega=2i K_\xi-i \int d^3x\,
\cP\frac{1}{\Delta} \d_i\pi^i,\label{eq:22}
\end{equation}
includes the contribution of the
longitudinal electric fields, and is explicitly given by 
\begin{equation}
  \cH^{\rm gs}=
 -\pi(A_0+\frac{1}{2\Delta}\pi)+\pi^0(\Delta
 A-\frac{\xi}{2}\pi^0)+i{\cal P}\rho- i \bar C\Delta\eta.\label{eq:22a}
\end{equation}

Turning on the chemical potential for electric charge can be done
through the shift $A_0(t,\vec x)\to A_0(t,\vec x)-\mu(\vec x)$ for a
time independent external source
$\mu(\vec x)$, since this changes $H^T\to H^T
+\int d^3x\, \mu(\vec x) \d_i\pi^i$ and thus to $H^T\to H^T
-\mu Q$ for constant $\mu$. 

In the case of a constant metric, supersymmetric
quantum mechanics is described by the action
\begin{equation}
  \label{eq:8}
  \begin{split}
    S^{\rm ss} =\int dt\ (iB_i\frac{d\phi^i}{dt}-i\bar\psi_i
    \frac{d\psi^i}{dt} + H^{\rm ss}),\\ H^{\rm ss}=\frac{\alpha}{2}
    g^{ij} B_i B_j +is \frac{\partial V}{\partial
      \phi^i}g^{ij}B_j-is\bar \psi_ig^{ij}\frac{\partial^2 V}{\partial
      \phi^j\partial \phi^k}\psi^k.
  \end{split}
\end{equation}
The entire action is BRST exact
\begin{equation}
S^{\rm ss}  = \int dt\,
    \Big\{\Omega,\bar\psi_i
    \big[i\frac{d\phi^i}{dt}+g^{ij}(\frac{\alpha}{2} B_j+is
    \frac{\partial V}{\partial \phi^j})\big]\Big\},
\end{equation}
where the BRST charge is 
\begin{equation}
\Omega=-i B_i\psi^i,\label{eq:26}
\end{equation}
and the fundamental Poisson brackets are
$\{\phi^i,B_j\}=-i\delta^i_j=-\{\psi^i,\bar\psi_j\}$, with all other
brackets vanishing. As consequence, the BRST transformations
${\rm s}=\{\Omega,\cdot\}$ are explicitly given by
\begin{equation}
{\rm s}\,\phi^i=\psi^i,\quad {\rm s}\,\psi^i=0,\quad {\rm s}\,
\bar\psi_i=B_i,\quad {\rm s}\, 
B_i=0. \label{eq:43}
\end{equation}
The Hamiltonian can be written as
\begin{equation}
  H^{\rm ss}=\half i\{\Omega,\bar\Omega\}, \quad \bar\Omega=-i\bar\psi_i
  g^{ij}(\alpha B_j+2is \frac{\partial 
    V}{\partial \phi^j}), \label{eq:40}
\end{equation}
with $\bar\Omega$ generating the so-called anti-BRST symmetry,
$\bar {\rm s}=\{\bar\Omega,\cdot\}$, explicitly given by
\begin{equation}
\bar {\rm s}\, \phi^i= \alpha g^{ij} \bar\psi_j,\ \bar {\rm s}\,
\psi^i= g^{ij}(\alpha B_j+2is \frac{\partial  
    V}{\partial \phi^j}),\
\bar {\rm s}\, \bar\psi_i= 0,\ \bar {\rm s}\, B_i= -2is\bar\psi_j
g^{jk}\frac{\partial^2  
  V}{\partial \phi^k\partial\phi^i}.\label{eq:44}
\end{equation}

The gauge sector can be written as a supersymmetric quantum mechanical
model with $H^{\rm gs}=-H^{ss}$ if $\alpha=s=1=\xi$, 
\begin{equation}
  \label{eq:41}
  \phi^i=\begin{pmatrix} A(\vec x) \\ A_0(\vec x)
  \end{pmatrix},\quad \psi^i= \begin{pmatrix} -\eta(\vec x) \\ i\rho
    (\vec x), 
  \end{pmatrix},\quad B_i=\begin{pmatrix} i\pi (\vec x) \\ -i\pi^0(\vec x)
  \end{pmatrix},\quad \bar\psi^i=\begin{pmatrix} i{\cal P}(\vec x) \\
    -\bar C (\vec x)
  \end{pmatrix},
\end{equation}
and 
\begin{equation}
  \label{eq:42}
  g^{ij}=\begin{pmatrix}-\frac{1}{\Delta}\delta^{3}(\vec x,\vec x') &
      0 \\ 0 & - \delta^{3}(\vec x,\vec x')
    \end{pmatrix},\quad V=\int d^3x A \Delta A_0, 
  \end{equation}
provided spatial integrations by parts are allowed. Formally, DeWitt's
condensed notation is used (in the sense that summation over $i$
includes an integration over $\vec x$, while $\delta^i_j$ includes a
Dirac delta function in three dimensions).

Such a reformulation is clearly not essential for an understanding of
the problem. Nevertheless, it indicates at this stage already that the
explicit computation of the partition function involves the value of
the exponential at the classical saddle point, the ``instanton''
solution $\frac{d\phi^i}{dt}=0$,
$\frac{\partial V}{\partial \phi^i}=0$.

\section{Planar vacuum capacitor}
\label{sec:plan-vacu-capac}

In this main section, the partition function for the vacuum capacitor
is computed, after identifying the complete Hilbert space from a
constrained Hamiltonian analysis that takes the non-trivial boundary
conditions of the physical set-up into account. Notations and
conventions are fixed in appendix \ref{sec:modes}. In order to
understand how the boundary conditions influence the result, it is
instructive to first review the standard and well-known results in the
case of periodic boundary conditions. This is done in appendix
\ref{sec:trivial} and \ref{sec:coher-stat-quart}, following
\cite{Henneaux:1992ig}.

\subsection{Spatial boundary conditions}
\label{sec:non-trivial-spatial}

For conducting plates, spatial boundary conditions on the fields have
to be imposed that implement
$\vec n \cdot \vec B=0=\vec n\times \vec E=0$ on the boundary. If
$x^i=(x^a,x^3)$ with $a=1,2$, this is guaranteed if the mode expansion
of $(A_a,\pi^a)$ contains sines only,
\begin{equation}
  \label{eq:49}
  A_c(x^i)=\sum_{n_a}\sum_{n_3>0} A^S_{c,k_a,k_3}\sin k_3 x^3 e^{ik_a
    x^a},
  \pi^d(x^i)=\sum_{n_a}\sum_{n_3>0} \pi^{Sd}_{k_a,k_3}\sin k_3 x^3 e^{ik_a
    x^a},
\end{equation}
with non-vanishing Poisson brackets
\begin{equation}
  \label{eq:71}
  \{A^S_{c,k_a,k_3},\pi^{*Sd}_{k'_a,k'_3}\}
  =\frac{2\delta^d_c}{V}\prod_{i=1}^{3}\delta_{n_i,n'_i}, 
\end{equation}
where $V=4L_1L_2L_3$. In order for bulk cancellations to work as in
the case of periodic boundary conditions, one is forced to use Neumann
conditions for $(A_3,\pi^3)$, so that 
\begin{equation}
  \label{eq:72}
  \begin{split}
  A_3(x^i)&=\sum_{n_a}[A^C_{3,k_a,0}+\sum_{n_3>0} A^C_{3,k_a,k_3}\cos k_3 x^3 ]e^{ik_a
    x^a},\\
  \pi^3(x^i)&=\sum_{n_a}[\pi^{C3}_{k_a,0}+\sum_{n_3>0}
  \pi^{C3}_{k_a,k_3}\cos k_3 x^3] e^{ik_a
    x^a}.
\end{split}
\end{equation}
This implies that 
\begin{equation}
  \label{eq:73}
  \{A^C_{3,k_a,0},\pi^{*C3}_{k'_a,0}\}
  =\frac{1}{V}\prod_{a=1}^{2}\delta_{n_a,n'_a},\quad
  \{A^C_{3,k_a,k_3},\pi^{*C3}_{k'_a,k'_3}\}
  =\frac{2}{V}\prod_{i=1}^{3}\delta_{n_i,n'_i},\ k_3>0.  
\end{equation}
These conditions are consistent with the boundary conditions used in
the context of the Casimir effect when one works in radiation gauge
$A_0=0$, $\d^i A_i=0$ (see e.g.~\cite{Ambjorn:1981xw}). The boundary
conditions on the remaining variables then follow from the Hamiltonian
analysis starting from 
$H_C=\int_B d^3x\, (\cH_0-A_0\d_i\pi^i)$. Indeed, in order to impose
the Gauss law in the bulk, $(A_0,\pi^0)$ should satisfy Dirichlet
conditions. In turn, the same then goes for the ghost pairs
$(\eta,\cP)$, $(\bar C,\rho)$, and also for $(A,\pi)$. Again, this is
consistent with the conditions in the context of the Casimir effect
(e.g.~\cite{Ambjorn:1982en} where it is shown that there is a standard
supersymmetric cancellation between the zero point energies of the
gauge sector, and also \cite{Vassilevich:1994cz},
\cite{Esposito:1997ja} \cite{Esposito:2000dy} for related
considerations).

\subsection{Degrees of freedom and dynamics}
\label{sec:hamiltonian-analysis}

When substituting the mode expansion, the canonical Hamiltonian splits
into three pieces,
\begin{equation}
  \label{eq:83}
  H_C=H_B+H_{W}+H_{NPG},
\end{equation}
with a standard bulk piece
\begin{align}
  \label{eq:82}
 & H_B= \frac{V}{4}\sum_{n_a,n_3>0}\Big[
   \pi^{bS}_{k_a,k_3}\pi^{*S}_{b,k_a,k_3}
  +\pi^{3C}_{k_a,k_3}\pi^{*3C}_{b,k_a,k_3}\\
  & +k_1^2(A^S_{2,k_a,k_3}A^{*S}_{2,k_a,k_3}+A^{C}_{3,k_a,k_3}A^{*C}_{3,k_a,k_3})
    +k_2^2(A^S_{1,k_a,k_3}A^{*S}_{1,k_a,k_3}+A^{C}_{3,k_a,k_3}A^{*C}_{3,k_a,k_3})
    \nonumber
  \\
  & +k_3^2(A^S_{1,k_a,k_3}A^{*S}_{1,k_a,k_3}+A^{S}_{2,k_a,k_3}A^{*S}_{2,k_a,k_3})
    +ik_2k_3(A^S_{2,k_a,k_3}A^{*C}_{3,k_a,k_3}-A^{C}_{3,k_a,k_3}A^{*S}_{2,k_a,k_3})
  \nonumber\\
  & +ik_1k_3(A^S_{1,k_a,k_3}A^{*C}_{3,k_a,k_3}-A^{C}_{3,k_a,k_3}A^{*S}_{1,k_a,k_3})
    -k_1k_2(A^S_{1,k_a,k_3}A^{*S}_{2,k_a,k_3}+A^{S}_{2,k_a,k_3}A^{*S}_{1,k_a,k_3})\Big].
    \nonumber 
\end{align}
The piece  
\begin{equation}
  \label{eq:24}
  H_W=\frac{V}{2}\sum_{n_a,n_3>0}\Big[A^S_{0,k_a,k_3}(ik_b\pi^{*S}_{b,k_a,k_3}+k_3
  \pi^{*3C}_{b,k_a,k_3})\Big],
\end{equation}
will give rise to the secondary constraints, $-ik_b\pi^{S}_{b,k_a,k_3}+k_3
  \pi^{3C}_{b,k_a,k_3}\approx 0$.
As expected and can be easily checked, there are no tertiary
constraints. 

The most interesting piece from the current perspective is 
\begin{equation}
  \label{eq:84}
  H_{NPG}=\frac{V}{2}\sum_{n_a}\Big[\pi^{3C}_{k_a,0}\pi^{3C*}_{k_a,0}
  +\omega^2_{k_a}A^{C}_{3,k_a,0}A^{C*}_{3,k_a,0}\Big],\quad
  \omega_{k_a}=\sqrt{k_1^2+k_2^2}.  
\end{equation}

In summary, we can split degrees of freedom according to whether they
are $k_3$ zero modes or not. In the latter group, we have
$(A_b,\pi^b)$, $(A_0,\pi^0)$, $(\eta,\cP)$, $(\bar C,\rho)$, which all
satisfy Dirichlet boundary conditions, as well as the $k_3\neq 0$
modes of $(A_3,\pi^3)$ satisfying Neumann conditions.

The former group contains $(A^C_{3,k_a,0},\pi^{3C}_{k_a,0})$,
respectively the fields
\begin{equation}
  \label{eq:86}
  \phi(x,y)=\sum_{n_a}A^C_{3,k_a,0}e^{ik_b x^b},\quad
  \pi(x,y)=\sum_{n_a}\pi^C_{3,k_a,0}e^{ik_b x^b}.
\end{equation}
None of these variables is involved in any of the constraints. They are thus
physical. Note that while the associated vector potential and electric
fields are formally longitudinal,
\begin{equation}
  \label{eq:85}
  A^{NPG}_i(x,y,0)=\delta_i^3  \phi
  =\d_i[z \phi], \quad \pi^i_{NPG}(x,y,0)=\delta^i_3 \pi =\d_i[z \pi],
\end{equation}
this is not really the case since $z$ is restricted to the closed
interval $[0,L_3]$. Note also that the Poisson brackets for these
variables given in \eqref{eq:73} and the Hamiltonian \eqref{eq:84},
which are encoded in the bulk first order action restricted to these
degrees of freedom, completely determine the Lagrangian action of a
massless scalar in (2+1) dimensions after integrating out the momenta,
\begin{equation}
  \label{eq:87}
  S^{NPG}=\frac{L_3}{2}\int dt\int_{-L_1}^{L_1}dx\int_{-L_2}^{L_2}dy\Big[ (\dot
  \phi)^2-\d_a \phi\d^a \phi\Big].  
\end{equation}
In this context, the electric charge operator, by analogy with the
discussion in section \ref{sec:classical-thermo}, is taken to be the
quantum version of the classical observable
\begin{equation}
  \label{eq:88}
  Q=-\pi^{3C}_{0,0,0}A=-\int_{-L_1}^{L_1}dx\int_{-L_2}^{L_2}dy\,\pi,
  \quad A=4L_1L_2.
\end{equation}
which Poisson commutes both with the complete Hamiltonian and
all constraints. 

\subsection{Partition function}
\label{sec:partition-function}

For the non-zero mode sector of the theory, one can then follow the
analysis of the periodic case (fix the gauge, choose suitable
variables). The difference is only that the modes involved are
restricted to $k_3>0$. Up to details related to the standard Casimir
effect (which will be addressed elsewhere), one finds that the
contribution to the partition function from this sector is the
standard black body result, equation \eqref{eq:4}.

For the new sector, we first consider the non-zero modes of the non
proper gauge degrees of freedom, $(A^C_{3,k_a,0},\pi^{3C}_{k_a,0})$,
with $k_a\neq 0$. For them, one defines standard oscillator
variables
\begin{equation}
  \label{eq:89}
  a_{k_a}=\sqrt{\frac{\omega_{k_a}V}{2}}(A^C_{3,k_a,0}
  +\frac{i}{\omega_{k_a}}\pi^{3C}_{k_a,0}), 
\end{equation}
so that
\begin{equation}
  \label{eq:90}
  \{a_{k_a},a^*_{k'_a}\}=-i\delta_{n_a,n'_a},\quad
  H'_{NPG}={\sum}'_{n_a}\omega_{k_a}a^*_{k_a}a_{k_a}. 
\end{equation}
The contribution to the partition function,
\begin{equation}
  \label{eq:91}
  {Z}'_{NPG}(\beta,\rho)={\rm Tr}\,e^{-\beta\rho \hat H'_{NPG}},
\end{equation}
is given by 
\begin{equation}
  \label{eq:92}
  \ln {Z}'_{NPG}(\beta,\rho)=-{\sum}^\prime_{n_a}\ln(1-e^{-\beta\rho\omega_{k_a}}). 
\end{equation}
The standard approximation then leads to 
\begin{equation}
  \label{eq:93}
  \ln {Z}'_{NPG}(\beta,\rho)=-\frac{A}{4\pi^2}\int dk_1dk_2 \ln
  (1-e^{-\beta\rho\sqrt{k_1^2+k_2^2}})=\frac{A}{2\pi}\zeta(3)(\beta\rho)^{-2}. 
\end{equation}

For the zero mode of the non-proper gauge degrees of freedom, the
variables 
$q=A^{C}_{3,0,0}\sqrt{V}$, $p=\pi^{3C}_{3,0,0}\sqrt{V}$, have
canonical commutation relations, while the Hamiltonian and electric
charge observable are given by
\begin{equation}
  \label{eq:94}
  H^0_{NPG}=\half p^2,\quad Q=-\sqrt{\frac{A}{L_3}}p. 
\end{equation}
It follows that the contribution to the partition function, 
\begin{equation}
  \label{eq:95}
  Z^0_{NPG}(\beta,\nu,\mu)={\rm Tr} e^{-\beta\nu\hat
    H^0_{NPG}+\beta\mu\hat Q},
\end{equation}
of this
free particle is 
  \begin{equation}
    \label{eq:96}
    \ln{Z^0_{NPG}(\beta,\nu,\mu)}= \ln{\Delta q}-\half \ln{(
    2\pi\beta\nu)}+ \frac{\beta\mu^2}{\nu}\frac{A}{2L_3},
  \end{equation}
  where $\Delta q$ denotes the divergent interval of integration over
  $q$, which should be dropped.

  The starting point Hamiltonian corresponds to $\rho=1=\nu$, so that
  the semi-classical contribution to the partition function discussed in section
  \ref{sec:classical-thermo} is recovered through the last term of
  equation \eqref{eq:96}.

\section{Discussion and perspectives}
\label{sec:disc-persp}

We have used a Hamiltonian approach here in order to keep track of the
various degrees of freedom and of their nature. It should be
possible to streamline these derivations by using finite temperature
Lagrangian path integral methods combined with techniques from
topological field theory and extend the considerations here to more
complicated non trivial boundary conditions than the ones we have
treated explicitly.

The non-trivial effect is a zero mode effect, like in the case of
Bose-Einstein condensation \cite{Kapusta:1981aa}. The difference is
however that in the latter both observables $\hat H$ and $\hat N$
involve the same degrees of freedom, whereas in the our case, the
physical Hamiltonian $\hat H$ and the electric charge $\hat Q$ involve
different degrees of freedom. The electromagnetic analog of the
semi-classical Bekenstein-Hawking contribution to the partition
function comes here from the zero mode of the non-proper gauge degrees
of freedom, which are themselves zero-modes from the bulk perspective.

Magnetic charge can be treated in the same way when using a magnetic
instead of an electric formulation. Both types of charges
simultaneously can be understood in a manifestly duality invariant
first order formulation \cite{Deser:1976iy} (see also
e.g.~\cite{Schwarz:1993vs}) which includes an additional quartet
\cite{Barnich:2007uu,Barnich:2008ts}.

The next, in principle straightforward, step is then to generalize the
result discussed here to the spherical vacuum capacitor. For
linearized gravity around flat space, one can easily adapt the result
of \cite{Barnich:2010bu} and understand the Schwarzschild solution as
a coherent state of unphysical gravitons. Generalizing the derivation
here should also be tractable and is the object of a follow-up
project. This is then what an observer at spatial infinity would
see. He would however not be able to distinguish between a black hole
and a star from that computation alone.

It would be interesting to fully explore the consequences of the
present computation, both from a theoretical and experimental
viewpoint. Also interesting would be to understand in detail from the
current perspective what happens in full-fledged QED, how to resum
contributions from the gauge sector and to get different charged
sectors in the electromagnetic case, and similarily, to go from a flat
to a black hole background in the gravitational case.

As we have tried to show in \cite{Barnich:2010bu} and with this
computation here, in order to deal consistently with charged sectors
or black holes in the operator formalism, computations are transparent
when all polarizations of the
four potential or of the metric are quantized in a non-unitary
Hilbert space. This is also implicitly the case in the Euclidean path
integral formulation when choosing real paths for the Euclidean
version of $A_0$, or for the shift vectors. Since most of the
questions on black hole entropy have little to do with
transverse-traceless variables but rather with variables from the
gauge sector, one might want to take this specific non-unitarity into
account when discussing paradoxes related to black hole physics.

\section*{Acknowledgements}
\label{sec:acknowledgements}

\addcontentsline{toc}{section}{Acknowledgments}

This work is supported by the F.R.S.-FNRS Belgium, convention FRFC
PDR T.1025.14 and convention IISN 4.4503.15.

Part of the work has been done at the Kavli Institute for Theoretical
Physics China during the program ``Quantum Gravity, Black Holes and
Strings 2014''. Another part has been completed while visiting the
Perimeter Institute for Theoretical Physics. Research at Perimeter
Institute is supported by the Government of Canada through the
Department of Innovation, Science and Economic Development and by the
Province of Ontario through the Ministry of Research and Innovation.

The author is grateful to C\'edric Troessaert, Hern\'an Gonz\'alez,
Marc Geiller, Laurent Freidel and Marc Henneaux for useful
discussions.

\appendix

\section{Mode expansions}
\label{sec:modes}

\subsection{Periodic boundary conditions}
\label{sec:periodic}

Consider first periodic boundary conditions in a box $B_P$ with sides of
lengths $2L_i$ and volume $V_P=8L_1L_2L_3$.  Note that in this case, no
improvement terms are needed for the gauge fixed Hamiltonian
$H_0+\{\Omega,K_\xi\}$. The fields
\begin{equation}
z^A=(A_0,\pi^0,A_i,\pi^i,\eta,\cP,\bar C,\rho)\label{eq:54}, 
\end{equation}
are expanded in terms of Fourier series at fixed time $t$,
\begin{equation}
  \label{eq:51}
  z^A(x^i)=\sum_{n_i} z^A_{k_i}e^{ik_ix^i},\quad  z^A_{k_i}= z^{A*}_{-k_i},\quad
   z^A_{k_i}=\frac{1}{{V_P}}\int_{B_P} d^3x\, z^A(x^i)e^{-ik_ix^i}, 
\end{equation}
with $n_i\in \mathbb Z$ and $k_i=\frac{\pi n_i}{L_{(i)}}$ (no
summation over $i$).
Quadratic integrals are related as
\begin{equation}
  \label{eq:47}
  \int_{B_P}d^3x\, z^A(x^i)z^B(x^i)= V_P \sum_{n_i}
  z^A_{k_i}z^{*B}_{k_i}. 
\end{equation}

The canonical Poisson bracket relations that
originate from the kinetic term
\begin{equation}
  \int_{B_P} d^3x\, \dot \phi(x^i,t)\pi(x^i,t),
\end{equation}
for each canonically conjugated pair are then
\begin{equation}
  \label{eq:52}
\{ z^A_{k_i}, z^{*B}_{k'_i}\}=\frac{\sigma^{AB}}{V_P}\prod_{i=1}^3
\delta_{n_i,n'_i}, 
\end{equation}
with all other Poisson brackets following from the middle of equation
\eqref{eq:51}. Here $\sigma^{AB}$ is the canonical symplectic matrix
obtained by combining
\begin{equation}
  \begin{pmatrix} 0 & 1\\
    -1 & 0
  \end{pmatrix}
\end{equation}
for each canonical pair. Translating back to position space gives 
\begin{equation}
  \label{eq:70}
  \{z^A(x^i),z^B(y^i)\}=\sigma^{AB}\delta^{(3)}_P(x^i,y^i),\quad
  \delta^{(3)}_P(x^i,y^i)=\frac{1}{V_P}\sum_{n_i}e^{ik_i (x^i-y^i)}. 
\end{equation}

Alternatively, if one replaces the exponentials by sines and cosines
in the $z=x^3$ direction,  
\begin{equation}
  \label{eq:46}
  z^A(x^i)=\sum_{n_a}\big[c^A_{k_a,0}+\sum_{n_3>0}(c^A_{k_a,k_3}
  \cos{k_3x^3}+s^A_{k_a,k_3}\sin{k_3
x^3})\big]e^{ik_ax^a}, 
\end{equation}
with $a=1,2$
\begin{equation}
  \label{eq:68}
  c^A_{k_a,0}=\frac{1}{V_P}\int_{B_P} d^3x\, z^A(x^i)e^{-ik_a x^a}= z^A_{k_a,0},
\end{equation}
and, for $k_3> 0$, 
\begin{equation}
  \begin{pmatrix} c^A_{k_a,k_3} \\ s^A_{k_a,k_3}
  \end{pmatrix}
  =\frac{2}{V_P} \int_{B_P} d^3x\, z^A(x^i)e^{-ik_a
    x^a}\begin{pmatrix} \cos{k_3x^3}\\ \sin{k_3x^3}
  \end{pmatrix}=\begin{pmatrix} z^A_{k_a,k_3}+z^{*A}_{k_a,k_3}\\
    i(z^A_{k_a,k_3}-z^{*A}_{k_a,k_3})
  \end{pmatrix}. \label{eq:68a}
\end{equation}
In this case,
\begin{multline}
  \label{eq:78}
  \int_{B_P} d^3x\,z^A(x^i)z^B(x^i)=
  V_P\sum_{n_a}\Big\{c^A_{k_a,0}c^{*B}_{k_a,0}\\+\half\sum_{n_3>0}\big[
  c^A_{k_a,k_3}c^{*B}_{k_a,k_3}+s^A_{k_a,k_3}s^{*B}_{k_a,k_3}
  +i(c^A_{k_a,k_3}s^{*B}_{k_a,k_3}
  -s^A_{k_a,k_3}c^{*B}_{k_a,k_3})\big]\Big\}.
  \end{multline}
and the Poisson brackets are 
\begin{equation}
  \label{eq:67}
  \{c^A_{k_a,0},c^{*B}_{k'_a,0}\}=\frac{\sigma^{AB}}{V_P}\prod_{a=1}^2
\delta_{n_a,n'_a},
\end{equation}
and, for $k_3,k'_3> 0$,
\begin{equation}
  \label{eq:69}
  \{c^A_{k_3,k_a},c^{*B}_{k'_3,k'_a}\}=\frac{2\sigma^{AB}}{V_P}\prod_{i=1}^3
\delta_{n_i,n'_i}=\{s^A_{k_3,k_a},s^{*B}_{k'_3,k'_a}\},
\end{equation}
and all other Poisson brackets vanishing. In these terms, the periodic
delta function can be written as 
\begin{equation}
  \label{eq:74}
  \begin{split}
  \delta^{(3)}_{P}(x^i,y^i) & =\frac{1}{V_P}\sum_{n_a}e^{ik_a x^a}[1+2\sum_{n_3>0}
  \cos{k_3(x^3-y^3)}]\\ & = \frac{1}{V_P}\sum_{n_a}e^{ik_a x^a}[1+2\sum_{n_3>0}
  \cos{k_3x^3}\cos{k y^3}+\sin{k_3x^3}\sin{k_3y^3}].
\end{split} 
\end{equation}

\subsection{Neumann/Dirichlet boundary conditions}
\label{sec:neum-dir}

Imposing Neumann or Dirichlet boundary conditions on an interval of
length $L_3$ in the $z=x^3$ direction can be achieved by extending the
function of $z\in [0,L_3]$ to an even respectively odd function of
$z\in [-L_3,L_3]$. This amounts to setting $s^A_{k_3,k_a}$
respectively $c^A_{k_a,k_3}$ in \eqref{eq:46} to zero, while keeping
the definitions of the remaining modes in \eqref{eq:68} and
\eqref{eq:68a} unchanged (see \cite{SheikhJabbari:1999xd} for an
interpretation in terms of second class constraints). These formulas
can then be expressed in terms of the real volume $V=4L_1L_2L_3$ of
the body $B$ by the substitution $V_P=2V$.  In the Neumann case, we
now have
\begin{equation}
  \label{eq:80}
  \int_{B} d^3x\,
  z^A(x^i)z^B(x^i)=V\sum_{n_a}\big[c^A_{k_a,0}c^{*B}_{k_a,0}
  +\frac{1}{2}\sum_{n_3>0}c^A_{k_a,k_3}c^{*B}_{k_a,k_3}\big],
\end{equation}
while for the Dirichlet case,
\begin{equation}
  \label{eq:81}
  \int_{B} d^3x\,
  z^A(x^i)z^B(x^i)=\frac{V}{2}\sum_{n_a,n_3>0}s^A_{k_a,k_3}s^{*B}_{k_a,k_3}.
\end{equation}

The canonical Poisson brackets now originate from kinetic terms of
the form 
\begin{equation}
  \label{eq:77}
  \int_{B}d^3x \, \dot
  \phi(x^i,t)\pi(x^i,t)=\int^{L_1}_{-L_1} dx\int^{L_2}_{-L_2} dy\int^{L_3}_0 dz\, \dot
  \phi(x^i,t)\pi(x^i,t), 
\end{equation}
which implies that the brackets of the remaining modes in
\eqref{eq:67}, \eqref{eq:69} are to be multiplied by $2$, or
equivalently, in these equations, $V_P$ needs to be replaced by $V$.
In position space, one needs to replace
$\delta^{(3)}_P(x^i,y^i)$ in the RHS of \eqref{eq:70} by
$\delta^{(2)}_P(x^a,y^a)\Delta_{\pm}(x^3,y^3)$, with the $+$
corresponding to the Neumann and the $-$ to the Dirichlet case, and
where (see e.g.~\cite{Hanson1976}, chapter 4)
\begin{equation}
  \label{eq:48}
  \Delta_{\pm}(x^3,y^3)=\delta_{2L_3}(x^3-y^3)\pm
  \delta_{2L_3}(x^3+y^3)=\frac{1}{2L_3} \sum_{n_3} (e^{i k_3(x^3-y^3)}\pm
  e^{i k_3(x^3+y^3)}),
\end{equation}
and also
\begin{equation}
  \label{eq:75}
  \begin{split}
  \Delta_{+}(x^3,y^3)& =\frac{1}{L_3}+\frac{2}{L_3}\sum_{n_3>0}\cos{k_3x^3}\cos{k_3y^3},\\
  \Delta_{-}(x^3,y^3)& =\frac{2}{L_3}\sum_{n_3>0}\sin{k_3x^3}\sin{k_3y^3}.
\end{split}
\end{equation}

\section{Partition function for periodic boundary conditions}
\label{sec:trivial}

When there is no electric potential at the surface of the body, no
global electric charge and no non-trivial boundary conditions, the
theory is quantized in such a way that the contribution to the
partition function from the unphysical bosonic degrees of freedom
$(A_0,\pi^0)$, $(A,\pi)$ cancels the one from the ghost degrees of
freedom $(\eta,\cal P)$, $(\bar C,\rho)$ so that only the physical
degrees of freedom $(A_i^T,\pi^i_T)$ contribute. Let us briefly review
these computations. As we are ultimately interested in infrared
effects, we keep the volume finite and work with Fourier series
including zero modes, instead of Fourier integrals.

\subsection{Non-zero modes}
\label{sec:non-zero-modes}

For periodic boundary conditions in a box $B_P$ of volume
$V_P=8L_1L_2L_3$, we can adapt the change of variables from section
19.1.6 of \cite{Henneaux:1992ig} to the case of Fourier series instead
of Fourier integrals. In this case, $k_i=\frac{\pi n_{(i)}}{L_{(i)}}$
and one defines
\begin{equation}
  \label{eq:5}
 A'_0={\sum_{\vec n}}'\frac{1}{\sqrt{2\omega_{\vec k}V_P}}[a_{0,\vec
   k}\,e^{i\vec k\cdot\vec x}+{\rm c.c.}],\quad
 \pi^{0'}=i{\sum_{\vec n}}'\sqrt{\frac{\omega_{\vec k}}{2V_P}}[(a_{3,\vec
   k}+a_{0,\vec k})\,e^{i\vec k\cdot\vec x}-{\rm c.c.}],
\end{equation}
\begin{multline}
A'_i={\sum_{\vec n}}'\frac{1}{\sqrt{2\omega_{\vec k}V_P}}[a_{m,\vec k}\,e^m_{i,\vec
   k}\, e^{i\vec k\cdot\vec x}+{\rm c.c.}],\\
  \pi^{i'}=-i{\sum_{\vec n}}'\sqrt{\frac{\omega_{\vec
        k}}{2V_P}}[(a_{m,\vec k}\,e^m_{i,\vec
   k}+a_{0,\vec
   k})\,e^{i\vec k\cdot\vec x}-{\rm c.c.}],
\end{multline}
\begin{equation}
  \eta'=-{\sum_{\vec n}}'\frac{1}{\sqrt{2\omega^3_{\vec k}V_P}}[c_{\vec
    k}\,e^{i\vec k\cdot\vec x}+{\rm c.c.}],
  \quad {\cal P}'=i{\sum_{\vec n}}'\sqrt{\frac{\omega^3_{\vec k}}{2V_P}}
  [\bar c_{\vec k}\,e^{i\vec k\cdot\vec x}+{\rm c.c.}],
  \label{eq:62}
\end{equation}
\begin{equation}
\bar C'=-i{\sum_{\vec n}}'\sqrt{\frac{\omega_{\vec k}}{2V_P}}
  [\bar c_{\vec k}\,e^{i\vec k\cdot\vec x}-{\rm c.c.}],  \quad 
  \rho'=-{\sum_{\vec n}}'\frac{1}{\sqrt{2\omega_{\vec k}V_P}}
  [c_{\vec k}\,e^{i\vec
    k\cdot\vec x}-{\rm c.c.}],\label{eq:63}
\end{equation}
so that
\begin{equation}
  \label{eq:60}
  A'=-i{\sum_{\vec n}}'\frac{1}{\sqrt{
   2\omega^3_{\vec k}V_P}}[a_{3,\vec k}\,e^{i\vec k\cdot\vec x}-{\rm c.c.}],\quad
\pi'={\sum_{\vec n}}'\sqrt{\frac{\omega_{\vec k}^3}{2V_P}}[(a_{3,\vec
  k}+a_{0,\vec k})\,e^{i\vec k\cdot\vec x}+{\rm c.c.}],
\end{equation}
where ${\sum_{\vec n}}'=\sum_{\vec n\neq \vec 0}$, and
$\omega_{\vec k}=\sqrt{\vec k\cdot \vec k}$, while
$\{e^m_{i,\vec k}\}$ is an orthonormal triad, the first two vectors
being transversal and the
third longitudinal, $k^ie^1_{i,\vec k}=0=e^2_{i,\vec k}$ and
$e^3_{i,\vec k}=\frac{k_i}{\omega_{\vec k}}$.

Finally, there is an additional change of
variables to null oscillators, 
\begin{equation}
a_{\vec k}=a_{3,\vec k}+a_{0,\vec
  k},\quad b_{\vec k}=\half (a_{3,\vec k}-a_{0,\vec k})\label{eq:64}. 
\end{equation}

For the non-zero modes, if $a_{a,\vec k}$, $a=1,2$ are the
transverse physical oscillators,  while $a^\alpha_{\Gamma,\vec k}$,
$\alpha=1,2$, $\Gamma=1,2$, are
the null oscillators of the unphysical sector, with
$a^1_{\Gamma,\vec k}=(a_{\vec k},b_{\vec k})$ bosonic and $a_{\Gamma,\vec
  k}^2=(c_{\vec k},\bar c_{\vec k})$ fermionic, the 
non-vanishing Poisson brackets are
\begin{equation}
  \label{eq:11a}
\{a_{a,\vec k},  a^*_{b,\vec
  k'}\}=-i\delta_{ab}\delta_{\vec n,\vec n'},\quad
\{a^\alpha_{\Gamma,\vec k}, a^{\beta\,*}_{\Delta,\vec
k'}\}=-i\eta_{\Gamma\Delta}\delta^{\alpha\beta}\delta_{\vec n,\vec n'},  
\end{equation}
where indices are lowered (and raised) with 
$\delta_{ab}$, $\delta_{\alpha\beta}$ and the indefinite metric
$\eta_{\Gamma\Delta}$ given by 
\begin{equation}
  \label{eq:12}
\eta_{\Gamma\Delta}=\begin{pmatrix} 0 & 1\\ 1 & 0
\end{pmatrix}.
\end{equation}

The canonical Poisson brackets of the fields $z^A$ are then equivalent
to these non-zero modes Poisson brackets and the zero mode brackets
\begin{equation}
\{A_{0,\vec 0},\pi^0_{\vec 0}\}=1=\{A_{i,\vec 0},\pi^i_{\vec
  0}\}=-\{\eta_{\vec 0},\cP_{\vec 0}\}=-\{\bar C_{\vec 0},\rho_{\vec
  0}\}\label{eq:15}. 
\end{equation}

Note that longitudinal fields $A=A',\pi=\pi'$ do not have zero modes,
so that the commutation relations for the modes in a box imply
$\{A(\vec x),\Pi(\vec y)\}=-[\delta^{(3)}(\vec x,\vec
y)-\frac{1}{V_P}]$. How zero modes for these fields may be
re-introduced is briefly discussed in the next section.

With a view towards a subsequent large volume limit and a passage from
Fourier series to integrals, zero modes are usually neglected. In this
case,
$\sum_{\vec n}\to \frac{V}{(2\pi)^3}\int\,d^3k$,
$\delta_{\vec n,\vec n'}\to \frac{(2\pi)^3}{V}\delta^{(3)}(\vec k,\vec
k')$. If discrete and continuous Fourier coefficients/oscillators are
related by $z^A_{\vec k}\to \frac{\sqrt V}{(2\pi)^{3/2}} z^A(\vec k)$,
$a_{A\vec k}\to \frac{\sqrt V}{(2\pi)^{3/2}} a_A(\vec k)$ for all
$a_{a\vec k}$, $a_{\gamma,\vec k}^\alpha$, sums over $\vec n$ may
simply be replaced by integrals over $\vec k$ and Kronecker by Dirac
deltas in the above expressions for the mode expansions of the fields,
the Poisson brackets and quadratic expressions like the Hamiltonian or
the BRST charge.

\subsection{Zero modes}
\label{sec:zeros}

The piece of the BRST gauge fixed Hamiltonian (in Feynman gauge
$\xi =1$) $H^1=H_0+\{\Omega,K_1\}$ involving the zero modes
$z^A_{\vec 0}$ is
$\cH^1_{\vec 0}=\half \pi^i_{\vec 0} \pi_{i,{\vec 0}}-\half
\pi^0_{\vec 0}\pi^0_{\vec 0}+i \cP_{\vec 0} \rho_{\vec 0}$. When
$( A_{0,{\vec 0}},\bar \pi^0_{\vec 0})$ are quantized as
anti-Hermitian operators and the zero-mode ghosts in the Schr\"odinger
representation (cf.~\cite{Henneaux:1992ig}, sections 15.3.2 and
15.4.4), and after limiting the bosonic zero-mode integrations to
intervals $\Delta A_{\mu,{\vec 0}}$, their contribution to the
partition function would be
\begin{equation}
  \label{eq:55}
  Z(\beta)=\prod_{\mu=0}^3\frac{\Delta  A_{\mu,{\vec 0}}}{\sqrt{2\pi
      \beta}}\times\beta\times Z'(\beta),  
\end{equation}
with $Z'(\beta)$ the partition function for the non-zero modes. Note
also that the piece of the BRST charge involving zero-modes is
${\it \Omega}_{\vec 0}=-i\pi^0_{\vec 0}\rho_{\vec 0}$.

We will proceed differently however and start the analysis from the
zero mode contribution to the classical Lagrangian
$L=-\frac{1}{4} \int d^3x\, F_{\mu\nu}F^{\mu\nu}$. Indeed,
$\cL_{\vec 0}[ A_{\mu,\vec 0}]=\half \dot{ A}_{i,\vec 0}\dot{
  A^i}_{\vec 0}$. There then is only the primary constraint
$ \pi^0_{\vec 0}\approx 0$, but no secondary constraint. Introducing
the zero-mode ghost pair $({\bar C}_{\vec 0},\rho_{\vec 0})$, the
associated BRST charge is ${\it\Omega}_{\vec 0}$ given above. If one
would like the theory to also include the zero modes of the other
ghost pair, $(\eta_{\vec 0},\cP_{\vec 0})$, one can do so by adding a
suitable non-minimal sector. This is done by considering the zero-mode
Lagrangian as a function of the spurious $\cP_{\vec 0}$,
$\cL_{\vec 0}=\cL_{\vec 0}[ A_{\mu,\vec 0},-\cP_{\vec 0}]$. There then
is an additional constraint $-\eta_{\vec 0}\approx 0$, for which one
introduces the ghost pair $(\pi_{\vec 0}, A_{\vec 0})$, unrelated to
components of $( A_{i,\vec 0}, \pi^i_{\vec 0})$. The BRST charge
including this non-minimal sector is then
\begin{equation}
{\it \Omega}_{\vec 0}=-(
\pi_{\vec 0}\eta_{\vec 0}+i\pi^0_{\vec 0}\rho_{\vec
  0}).\label{eq:56}
\end{equation}
Choosing as gauge fixing fermion
\begin{equation}
  \label{eq:57}
   \frac{1}{2}i{\it \bar\Omega}_{\vec 0}=i{\bar C}_{\vec 0}(-A_{\vec
    0}-\half  \pi^0_{\vec 0})+ \cP_{\vec 0}(
  A_{0,\vec 0}-\half \pi_{\vec 0}), 
\end{equation}
the BRST gauge fixed Hamiltonian is
$\cH_{\vec 0}=\cH^{\rm ph}_{\vec 0}+\cH^{\rm gs}_{\vec 0}$, with 
\begin{equation}
\cH^{\rm ph}_{\vec 0}=\frac{1}{2}  \pi^i_{\vec 0}\pi_{i,\vec
    0},\label{eq:61}
\end{equation}
and $\cH^{\rm gs}_{\vec 0}=-\frac{1}{2}i\{{\it \Omega}_{\vec 0},{\it
  \bar \Omega}_{\vec 0}\}$, which is explicitly given by 
\begin{equation}
  \label{eq:58}
   \cH^{\rm gs}_{\vec 0}=-\pi_{\vec 0}( A_{0,\vec
    0}-\half \pi_{\vec 0})- \pi^0_{\vec 0}
  ( A_{\vec 0}+\half\pi^0_{\vec
    0})+i\bar C_{\vec 0}\eta_{\vec 0}+i\cP_{\vec 0}\rho_{\vec 0}. 
\end{equation}
When proceeding in this way, the longitudinal fields $(A,\pi)$ will
also include zero modes. Integrating out momenta can be done
consistently including the zero modes. The same applies to the mode
expansion of \eqref{eq:2}, \eqref{eq:22} with the understanding that
$\Delta$ goes to $-1$ for zero modes. When defining new variables for
zero-modes as for the non-zero modes (without a sum and with
$\omega_{\vec 0}=1$ in \eqref{eq:5}, \eqref{eq:60}, \eqref{eq:62},
\eqref{eq:63}), and in \eqref{eq:64}, the Poisson brackets of the
unphysical sector in \eqref{eq:11a} also include these zero modes.

When quantizing the unphysical zero-mode pairs
\begin{equation}
( A_{0,\vec 0},\pi^0_{\vec 0}),\quad ( \pi_{\vec
  0}, A_{\vec 0}),\quad ( \eta_{\vec
  0},\cP_{\vec 0}),\quad ({\bar C}_{\vec
  0},\rho_{\vec 0})\label{eq:59}, 
\end{equation}
in the Dirac-Fock representation, their contribution to the partition
function cancels through the same mechanism, reviewed in
appendix \ref{sec:bulk-canc} below, as for the non-zero modes of the unphysical
sector. One then remains with the (infinite) contribution of three
bosonic free particles encoded in \eqref{eq:61}, whose contribution to
the partition function is
\begin{equation}
  \label{eq:65}
  Z(\beta)=\prod_{i=1}^3\frac{\Delta A_{i,\vec
      0}}{\sqrt{2\pi\beta}}\times Z'(\beta). 
\end{equation}

\subsection{Bulk cancellations}
\label{sec:bulk-canc}

When inserting the mode expansion reviewed above, the BRST charge is
given by
\begin{equation}
  \label{eq:13}
  \Omega=\sum_{\vec n}(c^*_{\vec k}a_{\vec k}+a^*_{\vec k}c_{\vec k}). 
\end{equation}
In
Feynman gauge $\xi=1$, the gauge fixed Hamiltonian
\begin{equation}
H^{1}:=H_0+\{\Omega,K_1\}=H^{\rm ph}+ H^{\rm gs},\label{eq:66}
\end{equation}
is given by
\begin{equation}
  \label{eq:10a}
   H^{\rm ph}=\half \pi^i_{\vec 0}\pi_{i,\vec 0}+{\sum}'_{\vec n}
  \omega_{\vec k}a^*_{a,\vec k}a^a_{\vec
    k},\quad H^{\rm gs}=\sum_{\vec n}\omega_{\vec k} a^{\alpha\,*}_{\Gamma,\vec k}
  a_{\alpha,\vec k}^\Gamma. 
\end{equation}
Here $\omega_{\vec k}=\sqrt{k_i
  k^i}$ for the non-zero modes, $\omega_{\vec
  0}=1$ for the zero modes of the unphysical sector, $a_{a,\vec
  k}$,
$a=1,2$ are the transverse oscillators of the physical sector, while
$a^\alpha_{\Gamma,\vec
  k}$ are the bosonic and fermionic null oscillators of the unphysical
sector, with non-vanishing (graded) commutation relations
\begin{equation}
  \label{eq:11}
  [\hat a_{a,\vec k}, \hat a^\dagger_{b,\vec
    k'}]=\delta_{ab}\delta_{\vec n,\vec n'},\quad [\hat
  a^\alpha_{\Gamma,\vec k},
  \hat a^{\beta\,\dagger}_{\Delta,\vec
    k'}]=\eta_{\Gamma\Delta}\delta^{\alpha\beta}\delta_{\vec n,\vec n'}, 
\end{equation}
where indices are lowered and raised with the appropriate
metrics $\delta_{ab}$, $\delta_{\alpha\beta}$, $\eta_{\Gamma\Delta}$
and their inverses. 

At this stage, the difference with the partition function for a
complex scalar field, and with Bose-Einstein condensation, appears
clearly: the observable for which we would like to introduce a
chemical potential involves different degrees of freedom than the ones
of the Hamiltonian. Furthermore, such a BRST Fock space quantization
guarantees that only the physical sector contributes.  Indeed, since
\begin{equation}
\hat H^1=\hat H^{\rm ph}+\frac{1}{2}[\hat\Omega,\hat{\bar\Omega}],\quad
\half \hat{\bar\Omega}=\sum_{\vec n}\omega_{\vec k} (\hat{\bar
  c}^\dagger_{\vec k}
\hat b_{\vec
    k}+\hat b^\dagger_{\vec k}
  \hat{\bar c}_{\vec k})\label{eq:45}, 
\end{equation}
it follows that
$e^{-\beta\hat H^1}=e^{-\beta \hat H^{\rm ph}}+[\hat \Omega,\hat M]$
for some operator $\hat M$. The trace to be used for the partition
function is the Lefschetz trace, for which the sum over diagonal
matrix elements is weighted by minus one to the power the ghost number
of the state. In the context of supersymmetric quantum mechanics, this
corresponds to computing the Witten index. The Lefschetz trace of BRST
exact operators vanishes, while for a BRST closed operator, it agrees
with the Lefschetz trace of the operator in cohomology. Hence, in the
current set-up, the trace reduces to the trace for the physical
Hamiltonian in the physical Hilbert space associated to transverse
photons,
\begin{equation}
  \label{eq:16}
  {\rm Tr}_W\,e^{-\beta \hat H_1}={\rm Tr}_{\rm ph}e^{-\beta\hat
    H^{\rm ph}}. 
\end{equation}

Alternatively, in the context of path integral quantization, it is
convenient to introduce a collective notation $a_A$ for all the
oscillators $a_a,a^\alpha_\Gamma$. BRST Fock quantization is
implemented by using the holomorphic representation with boundary
conditions that fix that creation operators at $t'$, $a^*_A(t')=a^*_A$
and destruction operators at $t$, $a_A(t)=a_A$, (see
e.g.~\cite{berezin34method,Faddeev1991}, and also
\cite{Itzykson:1980rh}, chapter 9, \cite{Henneaux:1992ig}, chapters
15, 16). In order to be able to turn on a chemical potential, we
consider the coupling to a source by using
\begin{equation}
  \label{eq:79}
  \cH^j_{\vec k}=\omega_{\vec k}a^*_{A,\vec k}a^A_{\vec
    k}-a^*_{A,\vec k} j^A_{\vec k}(\tau)-a_{A,\vec
    k} j^{* A}_{\vec k}(\tau).
\end{equation}
The path integral representation of the kernel
$U^j_{\vec k}(t',t)$ at fixed $\vec k$ of the evolution operator
$e^{i(t'-t)\cH^j_{\vec k}}$ is then given by
$U^j_{\vec k}(t',t)=e^{i S^j_{\vec k}}|_{\rm extr}$, where the
classical action to be used is the one that has a true extremum when
taking into account the boundary conditions
\begin{equation}
  \label{eq:18}
  S^j_{\vec k}=\int^{t'}_t d\tau \Big[\frac{1}{2i}\big(\dot
  a^*_{A,\vec k}a^A_{\vec k}
  -a^*_{A,\vec k}\dot a^A_{\vec k}\big)-{\cal H}^j_{\vec k}\Big] 
  +\frac{1}{2i}\Big[(a^*_{A,\vec k}a^A_{\vec k})(t')
  + (a^*_{A,\vec k} a^A_{\vec k})(t)\Big].
\end{equation}
When using that the appropriate extremum is 
\begin{equation}
  \label{eq:20}
\begin{split}
  & a^A_{\vec k}(\tau)=e^{-i\omega_{\vec k}(\tau-t)}a^A_{\vec k}+i\int^\tau_td\tau'
  j^A_{\vec k}(\tau')e^{-i\omega_{\vec k}(\tau-\tau')},\\ 
  & a^{*A}_{\vec k}(\tau)=e^{-i\omega_{\vec k}(t'-\tau)}
  a^{*A}_{\vec k}+i\int^{t'}_\tau
  d\tau' j^{*A}_{\vec k}(\tau')e^{-i\omega_{\vec k}(\tau'-\tau)},
\end{split}
\end{equation}
one finds
\begin{multline}
  \label{eq:19}
\ln{{U^j_{\vec k}(t',t)}}=a^{*}_{A,\vec k} a^A_{\vec k}
  e^{-i\omega_{\vec k}(t'-t)} \\ +i\int^{t'}_td\tau\, \Big[ a^{*}_{A,\vec k} 
  j^A_{\vec k}(\tau)e^{-i\omega_{\vec k}(t'-\tau)}+
  j^{*A}_{\vec k}(\tau)e^{-i\omega_{\vec k}(\tau-t)}a_{A,\vec k}\Big]-\\
  -\int^{t'}_td\tau\Big[\int_t^{t'} d\tau' 
  j^{*}_{A,\vec k}(\tau)  \theta(\tau-\tau')e^{-i\omega_{\vec
      k}(\tau-\tau')}j^A_{\vec k}(\tau')\Big].
\end{multline}
When using a time independent source $j^c$ and $t'-t=-i\beta$, this gives 
\begin{multline}
  \label{eq:21a}
 \ln{ U^{j^c}_{\vec k}(\beta)}=a^{*}_{A,\vec k} a^A_{\vec k} e^{-\beta
      \omega_{\vec k}}+(a^{*}_{A,\vec k}j^A_{\vec k}+j^{*}_{A,\vec k}
    a^A_{\vec k})\omega_{\vec k}^{-1}(1-e^{-\beta \omega_{\vec k}})+\\+
    j^{*}_{A,\vec k}j^A_{\vec k}[\omega_{\vec k}^{-1}\beta
    -\omega_{\vec k}^{-2}(1-e^{-\beta \omega_{\vec k}})]. 
\end{multline}
When evaluating the trace in the holomorphic representation, one
should split into physical and unphysical oscillators. For each
physical oscillators, there is a pre-factor of
$(1-e^{-\beta \omega_{\vec k}})^{-1}$ coming from an appropriate
change of variables. As explicitly recalled in appendix
\ref{sec:coher-stat-quart}, these pre-factors cancel for the
unphysical oscillators. This cancellation corresponds to the one
between the bosonic and fermionic determinants in supersymmetric
quantum mechanics. As a result,
\begin{equation}
  \label{eq:25}
   {\rm Tr}\, e^{-\beta \hat \cH^{jc}_{\vec k}}=\frac{1}{(1-e^{-\beta
       \omega_{\vec k}})^2}e^{\omega_{\vec k}^{-1}\beta j^*_{A,\vec k}
     j^A_{\vec k} }
\end{equation}
In the absence of sources, when integrating over all the modes and
discarding the infinite contribution of the zero modes of the physical
sector, one finds the standard black body result,
\begin{equation}
  \label{eq:4}
  \ln Z'(\beta)=-2\frac{V}{(2\pi)^3}\int d^3k \ln{(1-e^{-\beta
      \omega_{\vec k}})}=\frac{V\pi^2}{45 \beta^3}.   
\end{equation}
Note that, if instead of the kernel of the evolution operator, one
directly computes the trace, the alternating sign in the Lefschetz
trace is taken into account through periodic boundary conditions in
imaginary time for the ghosts (see e.g.~\cite{Bernard:1974bq} for
finite temperature QED or \cite{AlvarezGaume:1986nm} for
supersymmetric quantum mechanics), so that all fields satisfy periodic
boundary conditions in imaginary time.

Note also that, in real time, indefinite metric quantization is
implemented in the path integral through imaginary values for the
paths associated to $(A_0,\pi^0)$ (cf.~\cite{Henneaux:1992ig} page
355). In the Euclidean approach, when one substitutes $A_0$ by
$i\tilde A_0$ , these become then again real paths for $\tilde
A_0$. Conversely, this means that standard real paths for $\tilde A_0$
in the Euclidean approach correspond to using an indefinite metric
Hilbert space in real time.

Turning on a chemical potential for electric charge,
$H^{\rm gs}\to H^{\rm gs}+\mu Q$, with
$Q=-\sqrt{\frac{V}{2}}(a_{\vec 0}+a^*_{\vec 0})$ can be done in the
above computation through the coupling to the source,
\begin{equation}
j^A_{\vec
  k}=(j^a_{\vec k},j^\alpha_{\Gamma,\vec
  k})=(0,-\mu\sqrt{\frac{V}{2}}\delta^\alpha_1\delta_{\Gamma}^1\delta_{\vec
  n,\vec 0})\label{eq:76}
\end{equation}
and its complex conjugate. The result does not change:
$ e^{\omega_{\vec k}^{-1}\beta j^*_{A,\vec k} j^A_{\vec k} }=1$ due to
the metric $\eta_{\Gamma\Delta}$ used to contract indices of the
sources.

\section{Coherent states of quartets}
\label{sec:coher-stat-quart}

To a pair of bosonic null oscillators,
\[
[\hat a_\Gamma,\hat a^\dagger_\Delta]=\eta_{\Gamma\Delta},\quad
\eta_{\Gamma\Delta}=\begin{pmatrix} 0 & 1\\ 1 & 0
\end{pmatrix},
\]
one associates the coherent states, 
\[
|a\rangle= e^{ \hat a^\dagger_\Gamma a^\Gamma}|0\rangle,\quad 
\langle a^*|=\langle 0|e^{a^{*\Gamma}\hat a_\Gamma}.
\]
 Their overlap is given by 
\[
\langle a^*|a \rangle =e^{a^{*\Gamma}a_\Gamma},
\] 
while the completeness relation is 
\[
\hat 1=\int \prod_{\Gamma=1,2}\frac{d a^*_\Gamma da_\Gamma}{2\pi i} e^{-a^{*\Delta}
    a_\Delta} |a\rangle \langle a^*|,
\]
with fundamental integral 
\begin{equation}
I[j,j^*]=\int \prod_{\Gamma=1,2}\frac{d a^*_\Gamma da_\Gamma}{2\pi i} e^{-a^{*\Delta}
    a_\Delta  +a^{*\Delta} j_\Delta +j^{*\Delta}a_\Delta
  }=e^{j^{*\Delta}j_\Delta}\label{eq:7}. 
\end{equation}

Formulas for a pair of fermionic null oscillators, with anticommutation
relations given by $[\hat c_\Gamma,\hat
  c^\dagger_\Delta]=\eta_{\Gamma\Delta}$, are the same except
for the absence of $(2\pi i)^{-1}$ in the integration measure. 

Using the notation $a^\alpha_\Gamma=(a_\Gamma,c_\Gamma)$, for
$\alpha=1,2$, let $O(a^*;a)$ be the kernel of an operator $\hat O$ in
the Fock space of a quartet. In this representation, the Lefschetz
trace is given by
\[
 {\rm Tr}\, \hat O=\int \prod_{\alpha,\Gamma=1,2}\frac{d
    a^*_{\alpha,\Gamma} da_{\alpha,\Gamma}}{(2\pi i)^{2-\alpha}}
  O(a^*;a)
e^{-a^{*\Delta}_\gamma a^\gamma _\Delta}.
\]
For the operator $e^{-\beta\omega\hat N}$ with $\hat N=\hat
a^{*\Gamma}_\alpha \hat a^\alpha_\Gamma$ the counting operator for quartets,
the kernel is 
\[ 
\langle a^*| e^{-\beta\omega\hat N} |a \rangle =
e^{a^{*\mu}_\alpha a^\alpha_\mu e^{-\beta \omega}}, 
\]
so that
\[
{\rm Tr}\, e^{-\beta\omega\hat N}=\int \prod_{\alpha,\Gamma=1,2}\frac{d
    a^*_{\alpha,\Gamma} da_{\alpha,\Gamma}}{(2\pi i)^{2-\alpha}} 
e^{-a^{*\Delta}_\gamma a^\gamma_\Delta(1-e^{-\beta \omega})}=1.
\]
For the last equality, the change of variables
$a^{(*)}_{\alpha,\Gamma}\to a^{(*)}_{\alpha,\Gamma}(1-e^{-\beta
  \omega})^{-\half}$ leads to a vanishing Jacobian because
bosonic and fermionic contribution cancels, before using \eqref{eq:7}
with vanishing sources.

It also follows that 
\begin{equation}
  \label{eq:6}
  {\rm Tr}\, e^{-\beta\omega \hat N_b}=\frac{1}{(1-e^{-\beta\omega})^2}, 
\end{equation}
where $\hat N_b=\hat a^\dagger_\Gamma \hat a^\Gamma$ is the number operator for
the bosonic part of the quartet, i.e., for a pair of bosonic null
oscillators.


\providecommand{\href}[2]{#2}\begingroup\raggedright\endgroup

\end{document}